\documentclass[journal]{IEEEtran}
\usepackage[utf8]{inputenc}
\usepackage{color}
\usepackage{array}
\usepackage{verbatim}
\usepackage{float}
\usepackage{amsmath}
\usepackage{amsthm}
\usepackage{amssymb}
\usepackage{graphicx}
\usepackage{longtable}

\usepackage{verbatim}
\usepackage{float}
\usepackage{epstopdf}% To incorporate .eps illustrations using PDFLaTeX, etc.
\usepackage{subfigure}% Support for small, `sub' figures and tables
\usepackage{enumerate}
\usepackage{lmodern}
\usepackage{graphicx}
\usepackage{amsmath,mathrsfs,amssymb}
\usepackage{caption}
\usepackage{array}
\usepackage{longtable}
\usepackage{algpseudocode}
\usepackage{amsmath}
\usepackage{float}
\usepackage{algorithm}
\usepackage{booktabs}
\usepackage{array}

\usepackage[unicode=true,
 bookmarks=false,
 breaklinks=false,pdfborder={0 0 1},colorlinks=false]
 {hyperref}
\hypersetup{
 colorlinks,bookmarksopen,bookmarksnumbered,citecolor=blue,urlcolor=blue}
\usepackage[]{apacite}
\makeatletter
 \let\oldforeign@language\foreign@language
 \DeclareRobustCommand{\foreign@language}[1]{%
   \lowercase{\oldforeign@language{#1}}}

 \let\oldforeign@language\foreign@language
 \DeclareRobustCommand{\foreign@language}[1]{%
   \lowercase{\oldforeign@language{#1}}}

\ifCLASSINFOpdf
\else
\fi

\newcommand{\Lone}{$ \mathcal{L}_1 $}

\newtheorem{rem}{Remark}

\newtheorem{assum}{Assumption}

\begin{document}
%\fancyhf{}
%\lhead{\small{To cite this article:
%\bf{\textcolor{red}{H. A. Hashim, L. J. Brown, and K. McIsaac, “Nonlinear Stochastic Attitude Filters on the Special Orthogonal Group 3: Ito and Stratonovich,” IEEE Transactions on Systems, Man, and Cybernetics: Systems, vol. PP, no. -, pp. 1–13, 2018.}}}}
%\onecolumn
%
%Please note that where the full-text provided is the Author Accepted Manuscript or Post-Print version this may differ from the final Published version. \vspace{20pt} \\  { \bf To cite this publication, please use the final published version.} \vspace{10pt}\\
%{ \bf The published version (DOI) can be found at:  \href{http://dx.doi.org/10.1109/TSMC.2017.2702705}{10.1109/TSMC.2017.2702705} }
%
% \textbf{
% \begin{center}
% \vspace{20pt}
% Personal use of this material is permitted. Permission from the author(s) and/or copyright holder(s), must be obtained for all other uses, in any current or future media, including reprinting or republishing this material for advertising or promotional purposes.\vspace{60pt}\\
%\end{center}
% \begin{flushleft}
% \underline{Publication information:}\\
% Date of Submission: January 2018.\\
% Date of Acceptance: June 2018.\\
% Available Online: June 2018.\vspace{390pt}
%\end{flushleft} }
%\scriptsize{ \bf
%\vspace{20pt}Please contact us and provide details if you believe this document breaches copyrights. We will remove access to the work immediately and investigate your claim.
%} 
\onecolumn
\noindent\rule{18.1cm}{2pt}\\
\underline{To cite this article:}
{\bf{\textcolor{red}{H. A. Hashim, S. El-Ferik, and M. A. Abido., "A fuzzy logic feedback filter design tuned with PSO for L1 adaptive controller," Expert Systems with Applications, vol. 42, no. 23, pp. 9077-9085, 2015.}}}\\
\noindent\rule{18.1cm}{2pt}\\

{ \bf The published version (DOI) can be found at: \href{https://doi.org/10.1016/j.eswa.2015.08.026}{10.1016/j.eswa.2015.08.026} }\\

\vspace{40pt}Please note that where the full-text provided is the Author Accepted Manuscript or Post-Print version this may differ from the final Published version. { \bf To cite this publication, please use the final published version.}\\

\textbf{
	\begin{center}
		Personal use of this material is permitted. Permission from the author(s) and/or copyright holder(s), must be obtained for all other uses, in any current or future media, including reprinting or republishing this material for advertising or promotional purposes.\vspace{60pt}\\
	\end{center}
\vspace{360pt}
%	\begin{flushleft}
%		\underline{Publication information:}\\
% Date of Submission: December 2014.\\
%Date of Acceptance: September 2015.\\
%Available Online: February 2016.
%\end{flushleft}
 }
\footnotesize{ \bf
	\vspace{20pt}\noindent Please contact us and provide details if you believe this document breaches copyrights. We will remove access to the work immediately and investigate your claim.
} 

\normalsize

\twocolumn
\title{A Fuzzy Logic Feedback Filter Design Tuned with PSO for $ \mathcal{L}_1 $ Adaptive Controller}

\author{Hashim A. Hashim$^*$, Sami~El-Ferik, and~Mohamed~A.~Abido% <-this % stops a space
\thanks{$^*$Corresponding author, H. A. Hashim is with the Department of Systems Engineering, King Fahd University of Petroleum and Minerals, Dhahran, 31261, Saudi Arabia, e-mail: hmoham33@uwo.ca}% <-this % stops a space
\thanks{S.~El-Ferik is with the Department of Systems Engineering, King Fahd University of Petroleum and Minerals, Dhahran, 31261, Saudi Arabia.}% <-this % stops a space
\thanks{M. A. Abido is with the Department of Electrical Engineering, King Fahd University of Petroleum and Minerals, Dhahran, 31261, Saudi Arabia.}% <-this % stops a space
}

%\author{Hashim~A.~Hashim$^*$,~\IEEEmembership{~Member, IEEE}, Lyndon J. Brown, and~Kenneth McIsaac,~\IEEEmembership{~Fellow, IEEE}% <-this % stops a space
%\thanks{$^*$Corresponding author, H. A. Hashim, L. J. Brown and K. McIsaac are with the Department of Electrical and Computer Engineering,
%University of Western Ontario, London, ON, Canada, N6A-5B9, e-mail: hmoham33@uwo.ca, lbrown@uwo.ca and kmcisaac@uwo.ca.}}

\markboth{--,~Vol.~-, No.~-}{Hashim \MakeLowercase{\textit{et al.}}: A Fuzzy Logic Feedback Filter Design Tuned with PSO for $ \mathcal{L}_1 $ Adaptive Controller}
\markboth{}{Hashim \MakeLowercase{\textit{et al.}}: A Fuzzy Logic Feedback Filter Design Tuned with PSO for $ \mathcal{L}_1 $ Adaptive Controller}

\maketitle

\begin{abstract}
\Lone  adaptive controller has been recognized for having a structure that allows decoupling between robustness and adaption owing to the introduction of a low pass filter with adjustable gain in the feedback loop. The trade-off between performance, fast adaptation and robustness, is the main criteria when selecting the structure or the coefficients of the filter. Several off-line methods with varying levels of complexity exist to help finding bounds or initial values for these coefficients. Such values may require further refinement using trial-and-error procedures upon implementation. Subsequently, these approaches suggest that once implemented these values are kept fixed leading to  sub-optimal performance in both speed of adaptation and robustness. In this paper, a new practical approach based on fuzzy rules for online continuous tuning of these coefficients is proposed. The fuzzy controller is optimally tuned using Particle Swarm Optimization (PSO) taking into accounts both the tracking error and the controller output signal range. The simulation of several examples of systems with moderate to severe nonlinearities demonstrate that the proposed approach offers improved control performance when benchmarked to \Lone adaptive controller with fixed filter coefficients.
\end{abstract}


\IEEEpeerreviewmaketitle{}

\section{Introduction}

\IEEEPARstart{R}{eal} systems are in general nonlinear, suffer from the presence of uncertainties in either their dynamics, model structures or parameters, and are often subject to unknown external disturbances. Enhancing the response of such systems by designing a controller that provides fast adaption as well as robustness is appealing. Unfortunately, the speed of adaptation and robustness are competing requirements and a trade-off is sought during the final selection of the controller.  Over the past couple of years, \Lone adaptive controller has been proposed as the controller that provides a mean to reduce the intertwined relation between fast adaption. 

\begin{figure*}[ht]
	\centering
	\includegraphics[scale=0.8]{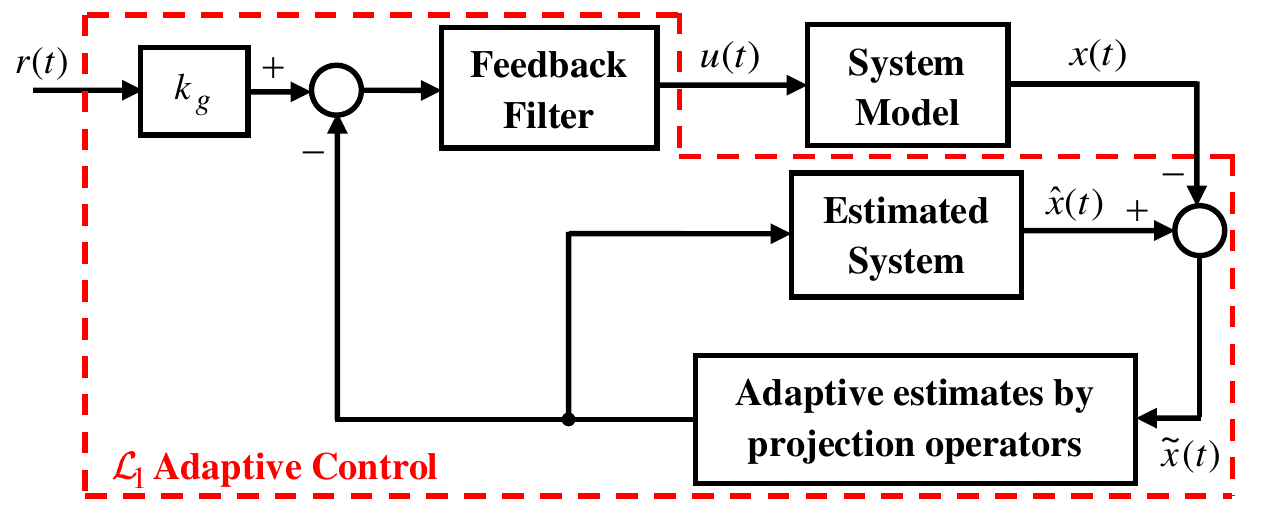}
	\caption{The general structure of L1 adaptive controller.}\label{Fuzzy_L1_Gen}
\end{figure*}

The structure of \Lone adaptive controller offers three features including the implementation of a low pass filter in order to limit the frequency range of the control signal and reduce the effect of the uncertainties (see Figure \ref{Fuzzy_L1_Gen}). 
The structure allows decoupling of the adaption and robustness using high-gain for fast adaption.
The filter is selected such that the system's output tracks properly the reference input and the undesirable uncertainties and frequencies are filtered ( see \cite{cao_design_2006} or \cite{hovakimyan_l1_2010}). Using the low pass filter, \Lone controller reduces the coupling between robustness and fast adaptation and provides infinity norm boundedness of the transient and steady state responses.
\Lone adaptive control was first introduced by \cite{cao_design_2006}.  It has been applied successfully to uncertain linear systems \cite{cao_design_2008}, uncertain nonlinear single-input-single-output (SISO) systems \cite{cao_guaranteed_2007}, \cite{luo_l_2010}, and nonlinear system multi-input-multi-output (MIMO) with unmatched uncertainties \cite{xargay_l1_2010-1}. And, the control approach showed satisfactory results on experimental flight tests \cite{gregory_l1_2009},  \cite{xargay_l1_2010}. 
The optimal structure of \Lone filter has been studied extensively in \cite{hovakimyan_l1_2010}. The trade-off between fast desired closed loop dynamics and filter parameters has been debated for long \cite{cao_design_2006,hovakimyan_l1_2010,li_filter_2008,li_optimization_2007,kharisov_limiting_2011,kim_multi-criteria_2014}.  Increasing the bandwidth of the low pass filter will reduce robustness margin, which will require slowing the desired closed loop performance in order to regain the robustness. However, slower selection of desired closed loop performance will deteriorate the output performance especially during the transient period \cite{hovakimyan_l1_2010}.  Limitations of \Lone adaptive controller and the interconnection between adaptive estimates and the feedback filter were studied in \cite{kharisov_limiting_2011}, where several filter designs were considered based on the use of disturbance observer. The authors showed that it is crucial to select the appropriate coefficients for a given filter to achieve the desired performance. Several attempts on identifying these optimal coefficients have been made in the literature.  This includes convex optimization based on linear matrix inequality (LMI) \cite{hovakimyan_l1_2010}, \cite{li_filter_2008} and multi-objective optimization using MATLAB optimization solver \cite{li_optimization_2007}. More recently, a systematic approach was presented in \cite{kim_multi-criteria_2014} to determine the optimal feedback filter coefficients in order to increase the zone of robustness margin. The authors proposed the use of greedy randomized algorithms. \\

\begin{figure*}[ht]
	\centering
	\includegraphics[scale=0.8]{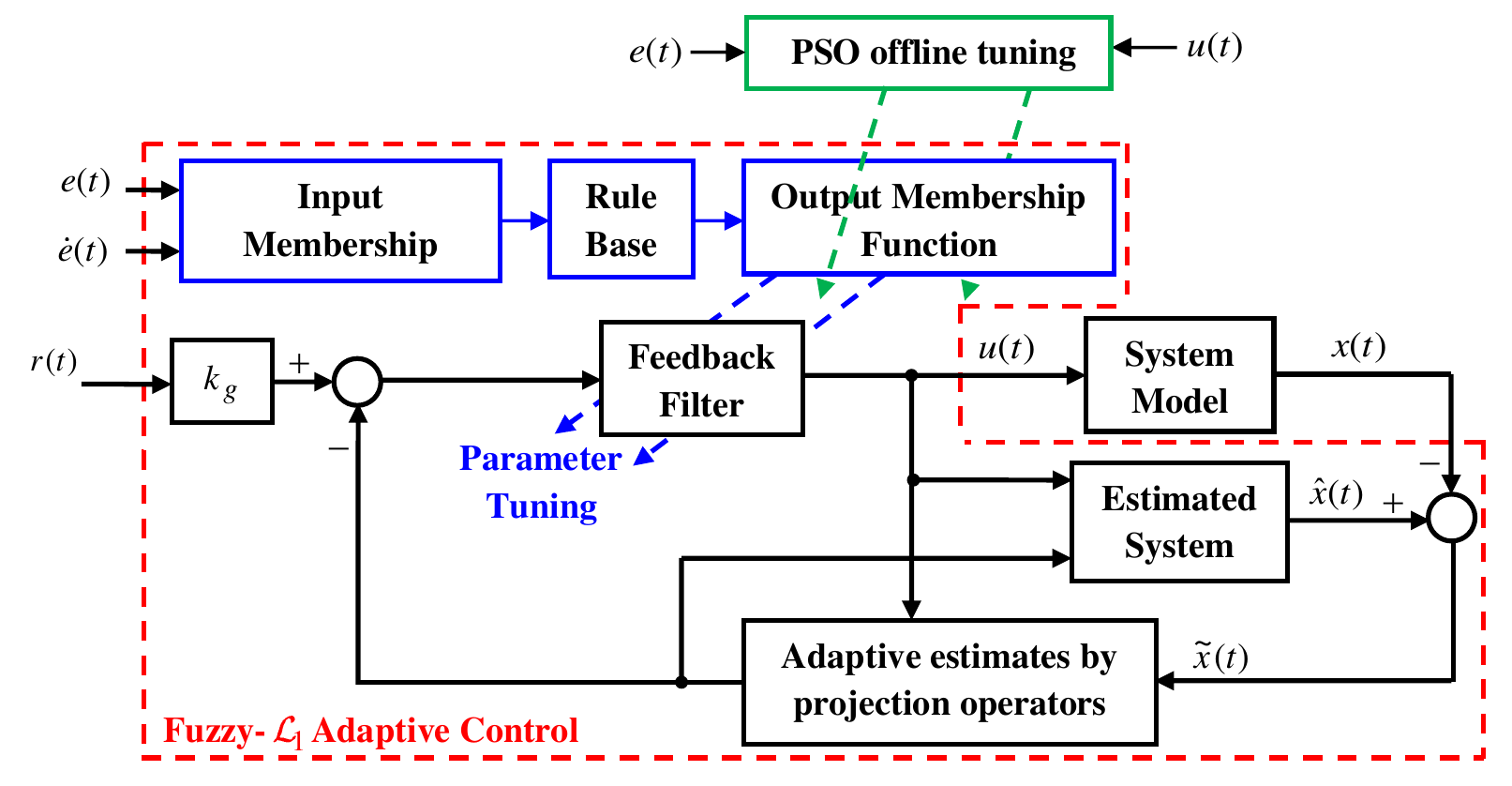}
	\caption{Proposed fuzzy- adaptive control structure.}
	\label{Fuzzy_L1_3PS}
\end{figure*}

One can observe that while the previous approaches to determine the optimal coefficients have different degrees of complexity, they agree on the fact that the selection of the appropriate coefficients is performed off-line; and once selected, these coefficients remain unchanged. This study claims that increasing the robustness while guaranteeing fast adaptation requires dynamic and on-line tuning of the feedback filter's coefficients and any proposed method should be relatively simple and easily implementable. To this end, this study proposes fuzzy tuning of the filter's coefficients optimized using PSO taking into account the rate and value of the tracking error between the model reference output and the system's output. The complete structure of fuzzy-\Lone adaptive controller is presented in Figure \ref{Fuzzy_L1_3PS}. The FLC-based tuning is performed on-line during operation. On the other hand, PSO identifies the optimal values of output membership functions through off-line tuning.
\\
Fuzzy logic controller (FLC) is classified as an intelligent technique and was first proposed in \cite{zadeh_fuzzy_1965}. FLC showed impressive results in control applications and it has been presented as a robustifying tool with adaptive controllers in \cite{tao_novel_2010},\cite{li_adaptive_2014}. It has been used to compensate unknown nonlinearities of twin rotor MIMO system with adaptive sliding mode control \cite{tao_novel_2010}.In \cite{li_adaptive_2014}, the authors suggested an observer-based adaptive backstepping control scheme and used FLC to approximate unknown uncertainties and to handle bounds of dead zone nonlinearity. On the other hand, evolutionary algorithms are introduced as potential optimization techniques in various control applications. They gained the interest of researchers and witnessed rapid developments over the past few decades. In particular, Particle swarm optimization (PSO) was introduced as a global search technique in \cite{eberhart_new_1995}. PSO has been applied successfully to optimize the structure and parameters of adaptive fuzzy controller in \cite{das_sharma_hybrid_2009} and optimize the variables of FLC membership functions in \cite{bingul_fuzzy_2011}, \cite{wong_pso-based_2008}. The need to tune controller systems with originally fixed coefficients has been widely recognized. In particular, fuzzy tuning has been investigated in several studies (see for instance \cite{precup2014novel}, \cite{valdez2014survey}, and \cite{kumar2014ann}) and controllers based on such approach have been implemented in many applications (see for instance \cite{kumar2014ann}, \cite{mendes2015indirect}, \cite{masumpoor2015adaptive}, \cite{zhang2015neuro}). This allows to conclude that the proposed approach is practical and can definitely be implemented with great benefit.\\

To summarize, in this work, fuzzy-\Lone adaptive controller is proposed to  tune the filter's coefficients in order to improve the trade-off between robustness and fast adaptation.  The coefficients are dynamically tuned and not kept fixed as in the literature, thus allowing for better performance. In the proposed approach, FLC is in charge of online tuning of the filter coefficients taking into account the range and rate of the tracking error. The use of FLC to tune the coefficients improves the stability and the robustness of the system and allows faster closed loop dynamics. Input membership functions and other FLC parameters are assigned arbitrarily while PSO optimizes the optimal variables of the fuzzy output membership functions. The approach is validated using different nonlinear systems and the extensive simulation results are benchmarked to the \Lone adaptive controller with fixed constant gain. The method is simpler than those in the literature and easily implementable.  \\

In order to guide the reader, the organization of the paper is as follows. In Section \ref{Sec_2}, a brief review of \Lone adaptive control including adaptation laws and the general structure is discussed.  Section \ref{Sec_3} presents the idea of filter design and the structure of the proposed controller. Section \ref{Sec_4} states the optimization problem and presents  the Particle Swarm Optimization (PSO) algorithm. Illustrative examples will be presented in Section \ref{Sec_5} in order to demonstrate the performance of the proposed approach. Finally, Section \ref{Sec_6} contains the concluding remarks and some suggestions for future work.

\section{Review of \Lone adaptive controller} \label{Sec_2}
{%\bf Problem formulation}\\
	Consider the following dynamics for nonlinear system
	\begin{equation}
	\label{eq:ch3L1sys}
	\begin{aligned}
	&\dot{x}(t) = A_{m}x(t) + b(\omega u(t) + f(x(t),u(t),t))\\
	& y(t) = cx(t)
	\end{aligned}
	\end{equation}
	
	where $x(t) \in \mathbb{R}^n$ is the system state vector (assumed measured); $u(t) \in \mathbb{R}$ is the control input; $y(t) \in \mathbb{R}$ is the system output; $b,c \in {R}^{n}$ are constant vectors (known);  $A_m $ is $\mathbb{R}^{n \times n}$ Hurwitz  matrix  (known) refers to the desired closed-loop dynamics; $\omega(t) \in \mathbb{R}$ is an unknown time variant parameter describes unmodeled input gain with known sign, and $f(x(t),u(t),t): \mathbb{R}^n\times \mathbb{R} \times \mathbb{R} \to \mathbb{R}$ is an unknown nonlinear continuous function.
	\begin{assum} 
		{\bf(Partially known with known sign control input)} Let the upper and the lower input gain bounds be defined by $\omega_l$ and $\omega_u$ respectively, where
		\begin{equation*}
		\omega \in \Omega \triangleq [\omega_l , \omega_u],\hspace{10pt} |\dot{\omega}| < \omega
		\end{equation*}
		$\Omega$ is assumed to be known convex compact set and $0<\omega_l<\omega_u$ are uniformly known conservative bounds.
	\end{assum}     
	
	\begin{assum} 
		{\bf (Uniform boundedness of $ f(0,u(t),t)) $} Let $B>0$ such that $ f(0,u(t),t)) \leq B $ for all  $ t \geq 0 $
	\end{assum} 
	\begin{assum} 
		{\bf (Partial derivatives are semiglobal uniform bounded)} For any $\delta >0$, there exist $d_{f_{x}} (\delta) >0 $ and $d_{f_{t}} (\delta) >0 $  such that for arbitrary $||x||_{\infty} \leq \delta$ and any $u$, the partial derivatives of $ f(x(t),u(t),t)) $ is piecewise-continuous and bounded,
		\begin{equation*}
		||\frac{\partial f(x(t),u(t),t) }{\partial x}|| \leq d_{f_{x}} (\delta),\hspace{10pt} |\frac{\partial f(x(t),u(t),t) }{\partial t}| \leq d_{f_{t}} (\delta)
		\end{equation*}  
	\end{assum} 
	\begin{assum} 
		{\bf (Asymptotically stable of initial conditions)} The system assumed to start initially with $x_0$ inside an arbitrarily known set $\rho_0$ i.e., $||x_0||_{\infty} \leq \rho_0 < \infty$.
	\end{assum}    
	\begin{equation}
	\label{eq:chfuzL1Max}
	\theta_b \triangleq d_{f_{x}}(\delta),\hspace{10pt} \Delta \triangleq B + \epsilon
	\end{equation}
	{\bf Lemma:} If $||x||_{\mathcal{L}_{\infty}} \leq \rho$ and there exist $u(\tau)$, $\omega(\tau)$, $\theta(\tau)$and  $\sigma(\tau)$ over $[0,t]$ such that
	\begin{equation}
	\omega_l<\omega<\omega_u
	\end{equation} 
	\begin{equation}
	\vspace{-2mm}
	|\theta(\tau)| < \theta_b
	\end{equation}
	\begin{equation}
	\vspace{-2mm}
	|\sigma(\tau)| < \sigma_b
	\end{equation}
	\begin{equation*}
	\vspace{-2mm}
	f(x(\tau),u(\tau),\tau) = \omega u(\tau) + \theta(\tau) ||x(\tau)||_{\infty} +  \sigma(\tau)
	\end{equation*}
	If $\dot{x}(\tau)$ and $\dot{u}(\tau)$ are bounded then $\omega(\tau)$, $\theta(\tau)$and $\sigma(\tau)$ are differentiable with finite derivatives.\\
	
	% {\bf $\mathcal{L}_1$ Adaptive Controller}\\
	
	The \Lone adaptive controller is composed of three parts defined as the state predictor, the adaption algorithm based on projection and the feedback filter (see Figure \ref{Fuzzy_L1_Gen}). 
	The main function of the state predictor is developed based on the adaptation laws
	\begin{equation}
	\label{eq:ch3L1est}
	\begin{aligned}
	&\dot{\hat{x}}(t) = A_{m}\hat{x}(t) + b(\hat{\omega} u(t) + \hat{\theta} ||x(t)||_{\infty}+\hat{\sigma} )\\
	& \hat{y}(t) = c\hat{x}(t)
	\end{aligned}
	\end{equation}
	The adaptive estimates $\hat{\omega} \in \mathbb{R}$, $\hat{\theta} \in \mathbb{R}$ and $\hat{\sigma} \in \mathbb{R}$ are defined as follows
	\begin{equation}
	\label{eq:chFuzL1Proj}
	\begin{aligned}
	&\dot{\hat{\omega}} = \Gamma Proj(\hat{\omega},-\tilde{x}^{\top}Pbu(t)), \hspace{10pt} \hat{\omega}(0) = \hat{\omega}_0  \\
	&\dot{\hat{\theta}} = \Gamma Proj(\hat{\theta},-\tilde{x}^{\top}Pb||x(t)||_{\infty}) \hspace{10pt} \hat{\theta}(0) = \hat{\theta}_0\\
	&\dot{\hat{\sigma}} = \Gamma Proj(\hat{\sigma},-\tilde{x}^{\top}Pb) \hspace{10pt} \hat{\sigma}(0) = \hat{\sigma}_0
	\end{aligned}
	\end{equation}
	where $\tilde{x} \triangleq \hat{x} - x(t)$, $\Gamma \in \mathbb{R}^{+}$ is the adaptation gain, and the solution of Lyapunov equation $A_m^TP+PA_m=-Q$ with symmetric  $P>0$ and $Q>0$. The projection operator ensures that $\hat{\omega} \in \Omega \triangleq [\omega_l , \omega_u]$, $\hat{\theta} \in \Theta \triangleq [-\theta_b,\theta_b]$, $|\hat{\sigma}| \leq \Delta$  with $\theta_b$ and $\Delta$ being defined in \eqref{eq:chfuzL1Max}. Projection operators will be evaluated as defined in \cite{pomet_adaptive_1992}\\
	\\
	%{\bf Control Law:} 
	With special interest to this paper, the control law is defined as 
	\begin{equation}
	u(s) = -k\,D(s)(\hat{\eta}(s)-k_g\,r(s))
	\end{equation}
	
	where $k > 0$ is a feedback gain and $D(s)$ is a strictly proper transfer function leading to a strictly proper and stable transfer function. The Laplace transforms of $r(t)$ and $\hat{\eta}(t) = \hat{\omega} u(t) + \hat{\theta} x(t)+\hat{\sigma}$ are $r(s)$ and $\hat{\eta}(s)$. Finally, $k_g$ is a necessary feedforward gain ensuring a unity steady state gain where $k_g \triangleq -1/(cA_{m}^{-1}b)$ ; $k > 0$. Thus, after a certain transient determined by its bandwidth, the effect of the filter will vanish from the dynamic of the closed loop system.\\
	Thus, in this case, the filter
	\begin{equation}
	C(s) = \frac{\omega \,k\,D(s)}{1+\omega \,k\,D(s)}
	\end{equation}       
	With DC gain $C(0) = 1$. The general structure of \Lone adaptive controller is depicted in Figure \ref{Fuzzy_L1_Gen}.
	\begin{rem}
		The main objective of this work is to design a FLC in order to tune the feedback gain of \Lone adaptive controller and ensure that $y(t)$ tracks a continuous reference signal $r(t)$. In addition, it is aimed at improving the robustness and tracking capability and reducing the control signal range when compared to \Lone adaptive controller with constant parameters.
	\end{rem}
	
	\section{Optimal Fuzzy-tuning of the feedback filter} \label{Sec_3}
	FLC has been used widely for various control applications. In this work, FLC is developed in order to tune the feedback filter gain of the \Lone adaptive controller. The importance of tuning this filter is crucial to improve the robustness and to reduce the control signal range.

	\subsection{Structure of Fuzzy Logic Controller}
	The error $e(t)$ is the difference between reference input $r(t)$ and regulated output $y(t)$. $k_p$ and $k_d$ are proportional and differential weights respectively. These parameters will be assigned before designing the membership functions and their values rely on the expected range of both $e(t)$ and $\dot{e}(t)$ in order to normalize fuzzy input between 1 and 0.
	\begin{equation}
	k_p \leq \frac{1}{||e||_{\infty}}, \hspace{10pt} k_d \leq \frac{1}{||\dot{e}||_{\infty}}
	\end{equation}  
	The existence of these norms is guaranteed by \Lone adaptive controller in case of stable dynamics. In addition, they can also be dynamically assigned. The fuzzy filter has a triangular membership functions for both inputs and output. The fuzzy filter has two inputs represented by the error and its rate and one output which is the inverse of the feedback gain $k_f$. Fuzzy inputs are the absolute values of $e(t)$ and $\dot{e}(t)$ multiplied by weighted gains $k_p$ and $k_d$. \Lone adaptive controller will consider the fuzzy output $k_f$ as a feedback gain if the error is greater than $k_e$. Adversely, the controller will consider a constant feedback gain $k$ if the error is less than or equal $k_e$ as shown in Figure \ref{Fuzzy_L1_4}.\\
	
	\section{Particle Swarm Optimization} \label{Sec_4}
	Particle swarm optimization is an intelligent evolutionary computation algorithm.  PSO algorithm deploys a set of particles in the space as a population and each particle is a candidate solution. Each particle in the search space moves randomly in swarm of particles to find the optimal solution. Each solution is defined by a particle position in the space and the velocity of swarming is necessary to target the best position. The proper setting of the algorithm variables ensures swarming in the vicinity space of the optimal solution and increases the probability of fast convergence. The velocity and position of the particle are defined according to the following two equations \eqref{eq:chFuzL1v} and \eqref{eq:chFuzL1x} respectively
	\begin{equation}
	\label{eq:chFuzL1v}
	\begin{split}
	v_{i,j}&(t) = \alpha(t)v_{i,j}(t-1) + c_{1}r_{1}(x_{i,j}^{*}(t-1) \\
	&-x_{i,j}(t-1))+ c_{2}r_{2}(x_{i,j}^{**}(t-1)-x_{i,j}(t-1))
	\end{split}
	\end{equation}
	\begin{equation}
	\label{eq:chFuzL1x}
	x_{i,j}(t) = v_{i,j}(t) + x_{i,j}(t-1))
	\end{equation}
	
	where $i=1,2,\cdots, N_p$ and $N_p$ is the population size, $j=1,2,\cdots, P_s$ and $P_s$ is the number of parameters In each particle, $x_{i,j}^{*}$ and  $x_{i,j}^{**}$ represent the local and global solutions respectively, $\alpha(t)$ is an exponential decreasing inertia, $c_1$ and $c_2$ represent personal and social influence of parameters and finally $r_1$ and $r_2$ are random numbers where $r_1 ,r_2 \in [0,1]$. The objective function is defined to enhance the tracking capability and improve the control signal range as follows
	\begin{equation}
	\label{eq:chFuzL1Obj}
	\begin{aligned}
	Obj = \sum\limits_{t=0}^{t_{sim}}\big(\gamma_1 e^2(t)+\gamma_2 u^2(t)\big)
	\end{aligned}
	\end{equation}
	where $e(t) = r(t) - y(t)$, $e(t)$ and $u(t)$ are the system error and control signal respectively. $\gamma_1$ and $\gamma_2$ are weights that can be selected arbitrarily. Obviously, the output membership functions have 18 parameters and they should be optimized to minimize the objective function. Particle swarm optimization is developed to search for the optimal values of aforementioned parameters. It must be noted that triangular nodes of output membership function represent position $x_{i,j}$, each two triangular intersect on the horizontal axis on one node. The computational flow diagram of PSO algorithm is illustrated in Figure \ref{Fuzzy_L1_PSO} \cite{abido_optimal_2002,mohamed2014improved}. The algorithm will be used with \Lone adaptive controller to define the optimal parameters of output membership functions for a specific number of generations as mentioned in \cite{eberhart_new_1995}, \cite{abido_optimal_2002}.
	\begin{figure}[!h]
		\centering
		\includegraphics[scale=0.6]{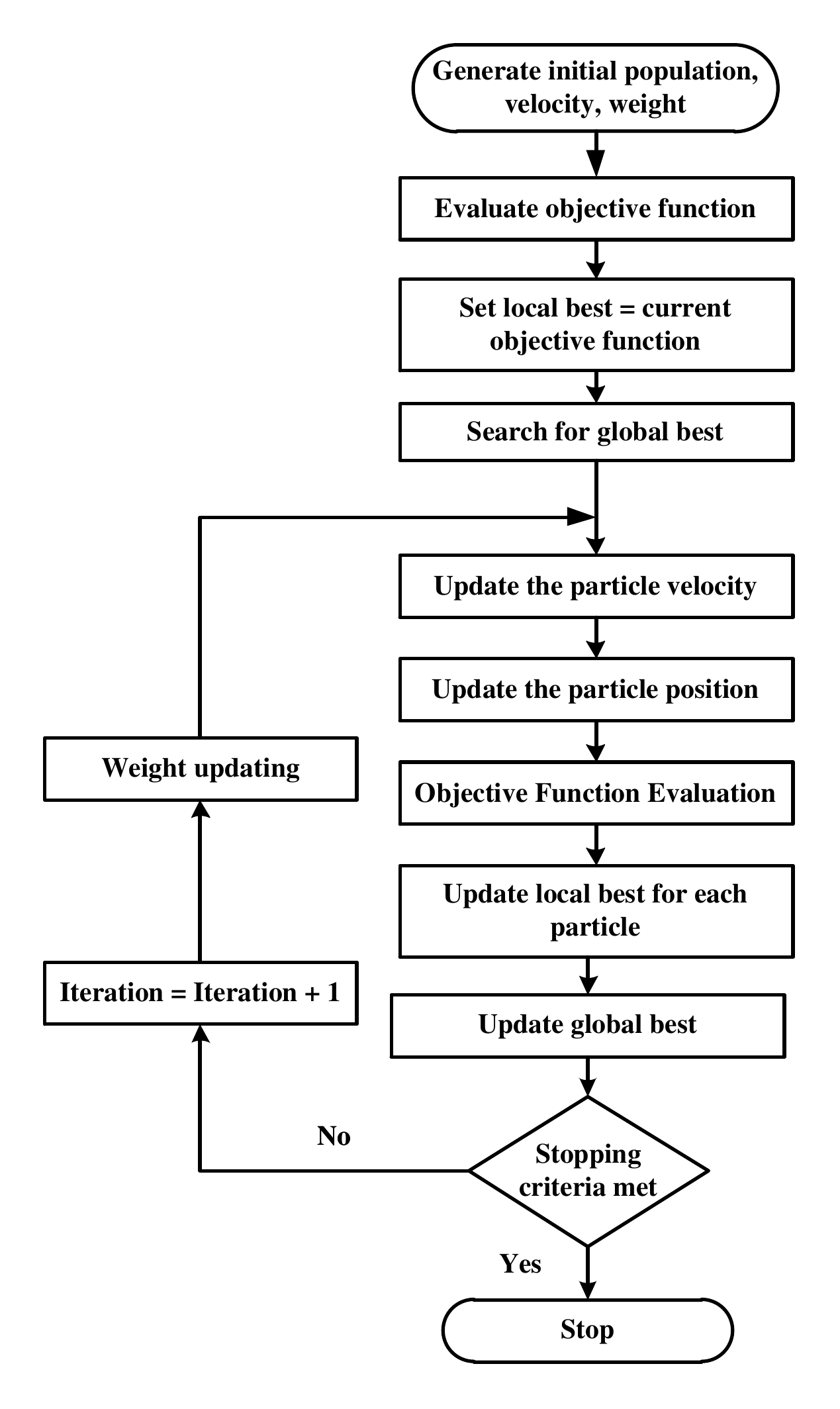}
		\caption{Flowchart of particle swarm Optimization.}\label{Fuzzy_L1_PSO}
	\end{figure}
	
	\begin{rem}
		In the proposed approach, the properties of the filter, such as strictly proper, low pass  with C(0)=1, are preserved. Consequently, stability of the Fuzzy-based-\Lone adaptive controller is guaranteed by the same analysis of stability done in \cite{cao_design_2006}.  
	\end{rem}       
	
	\section{Results and Discussions} \label{Sec_5}
	\subsection{Fuzzy \Lone adaptive controller implementation:}
	Problem in \cite{hovakimyan_l1_2010} has been considered here with additive nonlinearities added to the system as follows
	\begin{equation*}
	\begin{aligned}
	&\dot{x}(t) = A_{m}x(t) + B(\omega u(t) + f(x(t),t))\\
	& y(t) = Cx(t)
	\end{aligned}
	\end{equation*}
	where $x(t) = [x_1(t),x_2(t)]^{\top}$ are the system states, $u(t)$ is the control input, $y(t)$ is the regulated output and $f(t,x(t))$ includes high nonlinearity assumed to be unknown. In addition,
	\begin{equation*}
	A = 
	\begin{bmatrix}
	0 & 1\\
	0 & 0
	\end{bmatrix}, \hspace{10pt}
	B = 
	\begin{bmatrix}
	0\\
	1
	\end{bmatrix}, \hspace{10pt}
	C = 
	\begin{bmatrix}
	0 & 1
	\end{bmatrix}
	\end{equation*}
	and
	\begin{equation*}
	f(x(t),t) = 2x_1^2(t)+2x_2^2(t)+x_1sin(x_1^2)+x_2cos(x_2^2)\\
	\end{equation*}
	\begin{equation*}
	\omega = \frac{75}{s+75}\\    
	\end{equation*}
	$\omega$ is a function with fast dynamic to ensure smoothness of the control signal. The compact sets of the projection operators for unmodeled input parameters, uncertainties and disturbances were assigned to $[\omega_{min},\omega_{max}] \in [0,10]$, $\Delta = 100$ and $\theta_b = 10$ . The control objective is to design a fuzzy-\Lone adaptive controller to enhance each of control signal range and tracking capability of a bounded reference input $r(t)$ for the output signal $y(t)$. Desired poles are set to = $-21 \pm j0.743$, the constant feedback gain($k$) = 20, the adaptation gain($\gamma$) = 1000000 and $Q = \big[ \begin{smallmatrix} 1 & 0 \\ 0 & 1 \end{smallmatrix}\big]$. Fuzzy control parameters are $k_p = 0.1$ , $k_d = 0.05$ and $k_e = 0.1$. Figure \ref{Fuzzy_L1_4} illustrates the FLC with \Lone adaptive controller.
	\begin{figure*}[ht]
		\centering
		\includegraphics[width=12cm, height=7cm]{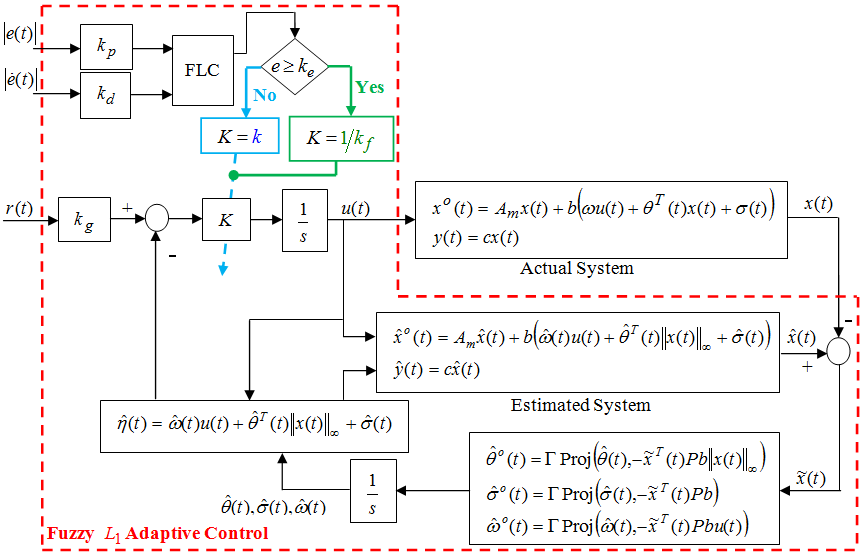}
		\caption{Fuzzy-\Lone adaptive controller for nonlinear SISO system.}\label{Fuzzy_L1_4}
	\end{figure*}

	\subsection{Membership Function Optimization}
	The objective of this work is to construct output membership function for FLC capable of reducing the error and the control signal. Values of input membership functions and constraints of the output membership functions were chosen based on trying different values by running a certain number of experiments. The range of input membership functions was adjusted between 0.08 and 1 and their values were selected as shown in Figure \ref{Fuzzy_L1_mem_e}. The fuzzy inputs and output have triangular membership functions with five linguistic variables. Linguistic variables are assigned as very large ($VL$), large ($L$), small ($S$), very small ($VS$) and zero ($Z$) where values of input membership function will be assigned arbitrarily. Values of output membership functions are  optimized using PSO. Rule base of the proposed filter is demonstrated in Table \ref{table:Tab_Rule}.
	\begin{table}[!t]
		\caption{Rule base of FLC.} % title of Table
		\centering % used for centering table
		\small
		\begin{tabular}{|c| c| c| c| c| c|} % centered columns (3 columns)
			\hline\hline %inserts double horizontal lines
			{\bf $\Delta e/e$} & {\bf VL}  & {\bf L} & {\bf S} & {\bf VS} & {\bf Z} \\ [0.0ex]
			%heading
			\hline\hline %inserts double horizontal lines
			{\bf VL} & $VL$ & $VL$ & $VL$ & $VL$ & $L$ \\[0ex]
			\hline                   
			{\bf L}  & $VL$ & $VL$ & $VL$ & $L$ & $S$ \\[0ex] 
			\hline
			{\bf S}  & $VL$ & $VL$ & $L$ & $S$ & $VS$ \\[0ex] 
			\hline
			{\bf VS} & $VL$ & $L$ & $S$ & $VS$ & $VS$ \\[0ex]
			\hline
			{\bf Z}  & $L$ & $S$ & $VS$ & $VS$ & $Z$ \\[0ex] 
			\hline\hline %inserts double horizontal lines
		\end{tabular}
		\label{table:Tab_Rule}
	\end{table}
	
	\begin{figure}[!h]
		\centering
		\includegraphics[scale=0.3]{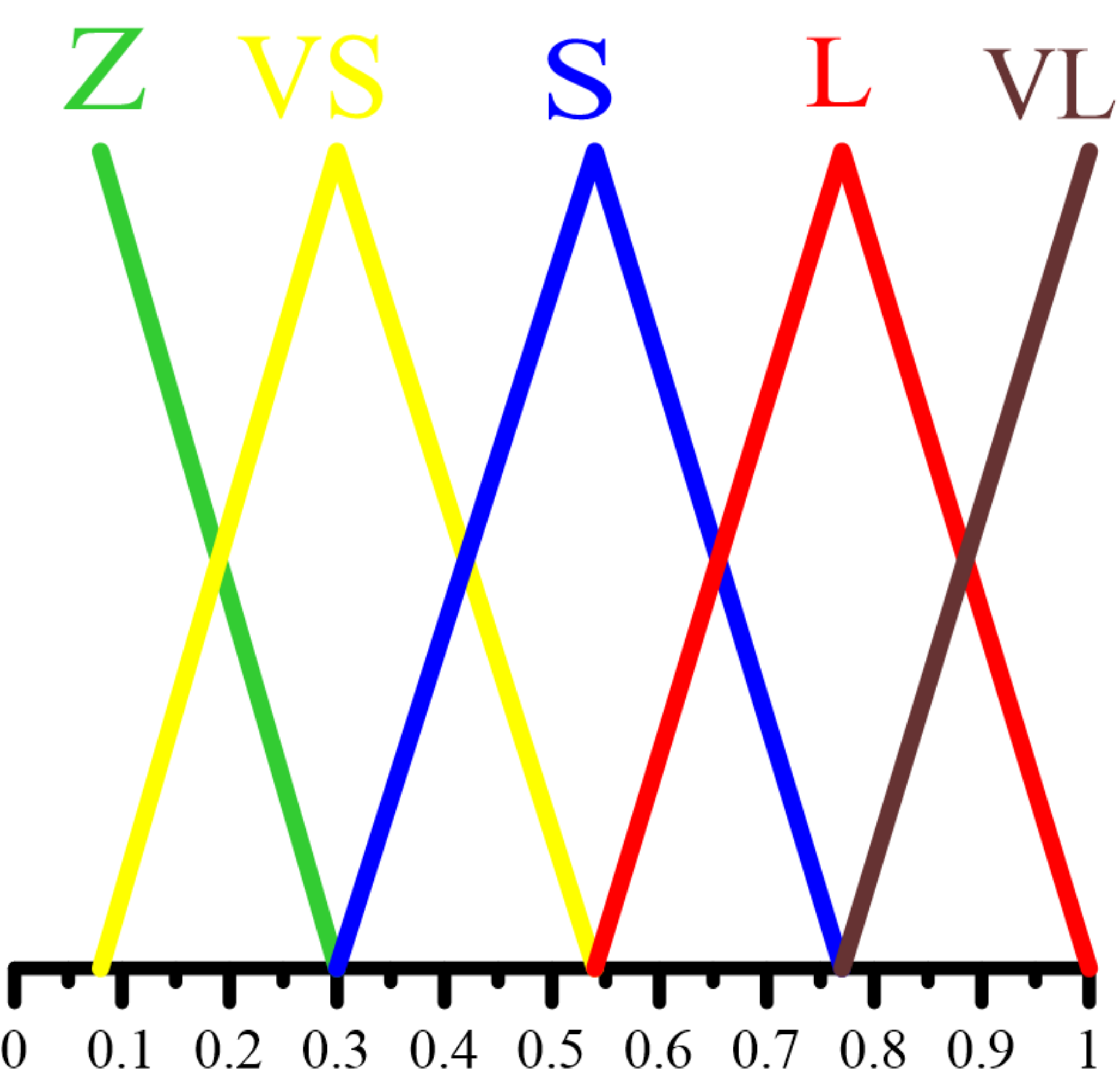}
		\caption{Error and rate of error membership functions.}\label{Fuzzy_L1_mem_e}
	\end{figure}
	
	Constraint values of output membership functions are represented by three parameters as lower ($l$), center ($c$) and higher ($h$) values. These three parameters of each triangular membership function will constrain between minimum and maximum bounds. Constraints bounds of the problem can be defined as follows 
	\begin{equation}
	\begin{aligned}
	&[4,8,8] \leq [VL_l,VL_c,VL_h] \leq [8,12,12]\\
	&[1.5,3,6] \leq [L_l,L_c,L_h] \leq [3,6,10]\\
	&[0.3,1.5,4] \leq [S_l,S_c,S_h] \leq [1.5,4,8]\\ 
	&[0,0.5,1.5] \leq [VS_l,VS_c,VS_h] \leq [0.5,1.5,3]\\  
	&[0.0,0.0,0.3] \leq [Z_l,Z_c,Z_h] \leq [0.0,0.0,1.5]\\ 
	\end{aligned}
	\end{equation}
	With $VL$, $ML$, $L$, $S$, $MS$, $VS$ and $Z$ were mentioned before as a linguistic variables. Also, we assigned $VL_c = VL_h$, $VL_l = S_h$, $L_l = VS_h$, $S_l = Z_h$, $VS_l = z_c$ and  $z_c = Z_l$ which means that we have only nine parameters to be optimized.
	\subsection{PSO Simulation results}
	The population size is set arbitrarily as 150 particles and each particle include 9 parameters will be optimized based on a minimization objective function and these parameter are $VL_c$, $VL_l$, $L_l$, $L_c$, $L_h$, $S_l$, $S_c$, $VS_l$ and $VS_c$ in (\ref{eq:chFuzL1Obj}). The initial settings of PSO algorithm are demonstrated in Table \ref{table:Tab_PSO} and the maximum numbers of generations is 100.
	\begin{table}[!h]
		\setlength{\tabcolsep}{5pt}
		\caption{Parameters setting for PSO.} % title of Table
		\centering % used for centering table
		\small
		\begin{tabular}{|c| c| c| c| c| c|} % centered columns (3 columns)
			\hline\hline %inserts double horizontal lines
			{\bf Parameter} & $\lambda$  & $\alpha$ & $c_1$ & $c_2$ \\ [0.0ex]
			%heading
			\hline\hline %inserts double horizontal lines
			{\bf Settings} & 10 & 0.99 & 2 & 2  \\[0ex]
			\hline\hline %inserts double horizontal lines
		\end{tabular}
		\label{table:Tab_PSO}
	\end{table}
	
	\subsection{PSO Results}
	The system was simulated for 8 seconds and the data was recorded every 0.01 seconds. The reference input was defined by $cos(0.5t)$ with zero initial conditions. The optimal variables of output triangular membership functions are illustrated in Figure \ref{Fuzzy_memb_u}. The fitness reduction during the search process is demonstrated in Figure \ref{Fuzzy_L1_PSOObj}. However, it is clear that objective function is reduced significantly and enormously to a suitable value which is reflected on the output performance as revealed in Figure \ref{Fuzzy_L1_out1}. Figure \ref{Fuzzy_L1_out1}.(a) demonstrates the optimal output performance and Figure \ref{Fuzzy_L1_out1}.(b) shows the control signal of the considered problem.\\
	\begin{figure}[!h]
		\centering
		\includegraphics[scale=0.3]{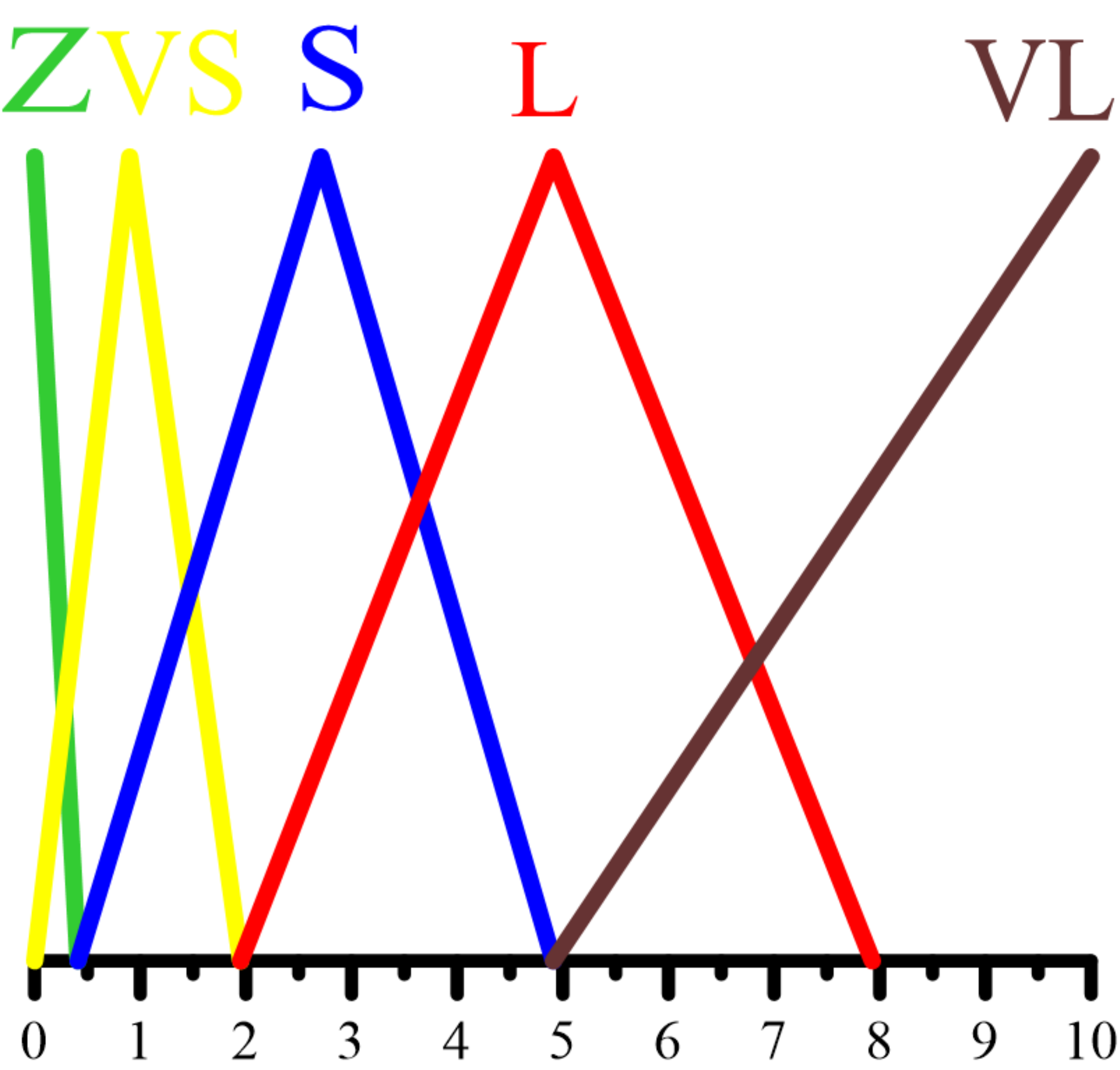}
		\caption{Graphical illustration of output membership functions.}\label{Fuzzy_memb_u}
	\end{figure}
	\begin{figure*}[ht]
		\centering
		\includegraphics[scale=0.6]{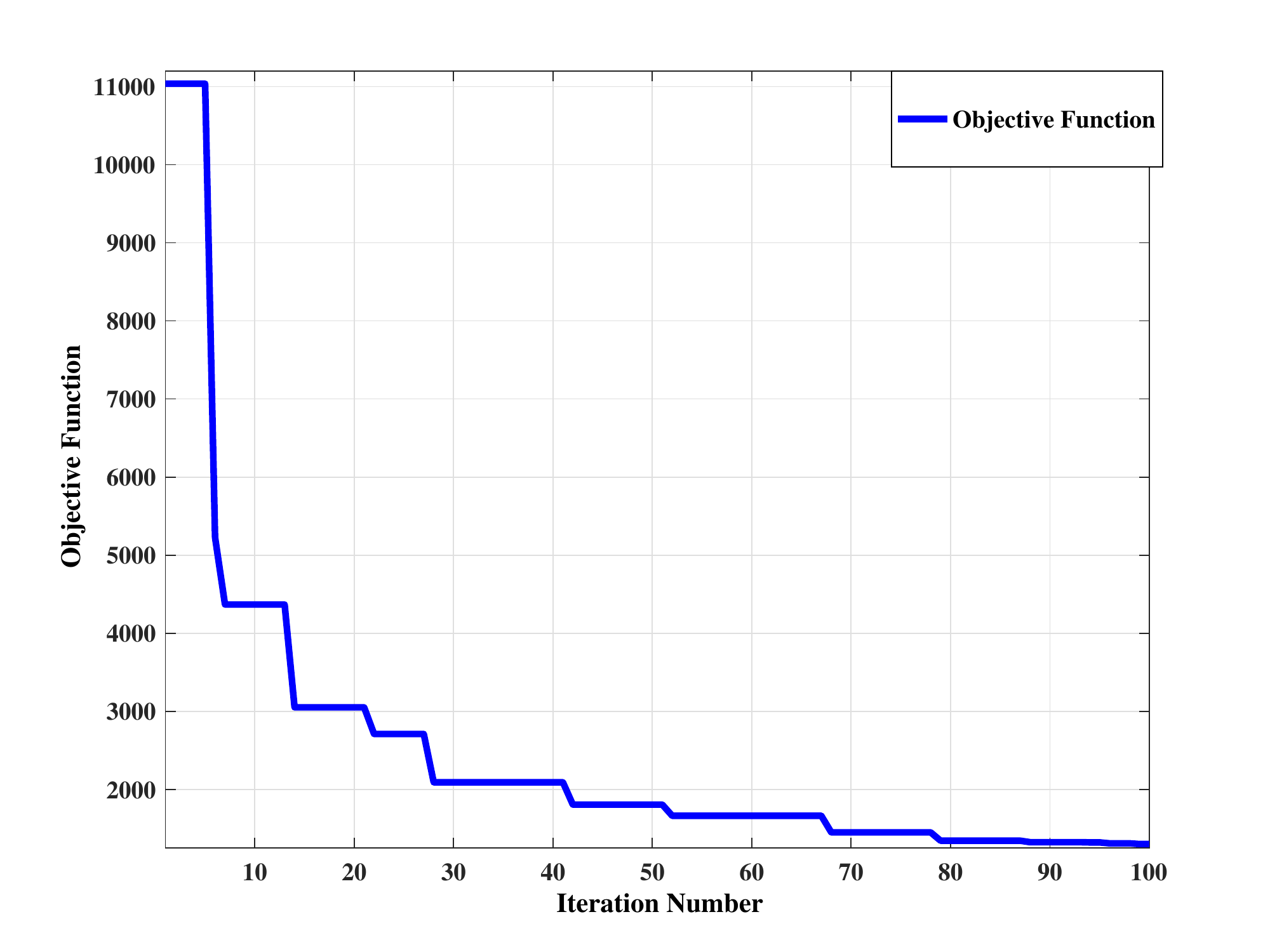}
		\caption{Objective function minimization with PSO search process.}\label{Fuzzy_L1_PSOObj}
	\end{figure*}
	\begin{figure*}[ht]
		\centering
		\includegraphics[scale=0.6]{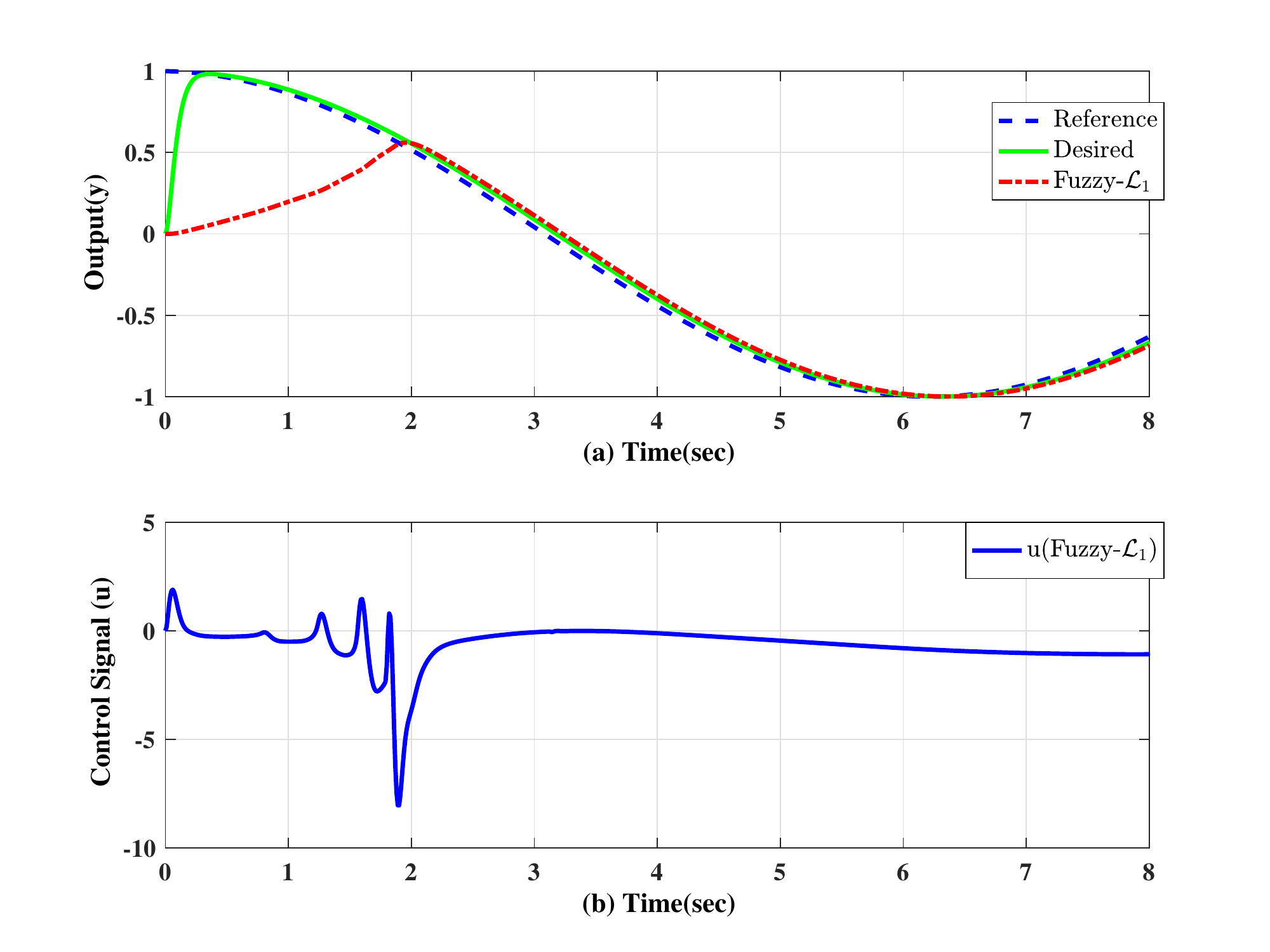}
		\caption{Performance of fuzzy-\Lone adaptive controller after 100 iterations search process.}\label{Fuzzy_L1_out1}
	\end{figure*}
	In this study, three different scenarios are considered to demonstrate the robustness of fuzzy-\Lone adaptive controller. All cases will be simulated for 40 seconds.
	The first case will discuss the nonlinear system included in the search process. Case 2 includes the nonlinear model with high uncertainties, unmodeled input parameters and adding some disturbances in order to validate the robustness of fuzzy filter with \Lone adaptive controller. Case 3 consider all assumptions in case 2 in addition to investigate the system with faster desired closed loop dynamics.\\
	{\bf Case 1:} Figure \ref{Fuzzy_L1_out2} presents the output performance of fuzzy-\Lone adaptive controller versus \Lone adaptive controller and their control signals. Fuzzy-\Lone adaptive controller guarantees uniform transient and smooth tracking performance. In addition, its major contribution lies in reducing the control signal range by tuning the feedback gain. Tuning feedback gain enhances the robustness of the system and reduces the control signal range. The correspondence difference of feedback gain between fuzzy-\Lone adaptive controller and \Lone adaptive controller is illustrated in Figure \ref{Fuzzy_L1_out2_2}.(a). The errors of both controllers are presented in Figure \ref{Fuzzy_L1_out2_2}.(b).
	\begin{figure*}[ht]
		\centering
		\includegraphics[scale=0.6]{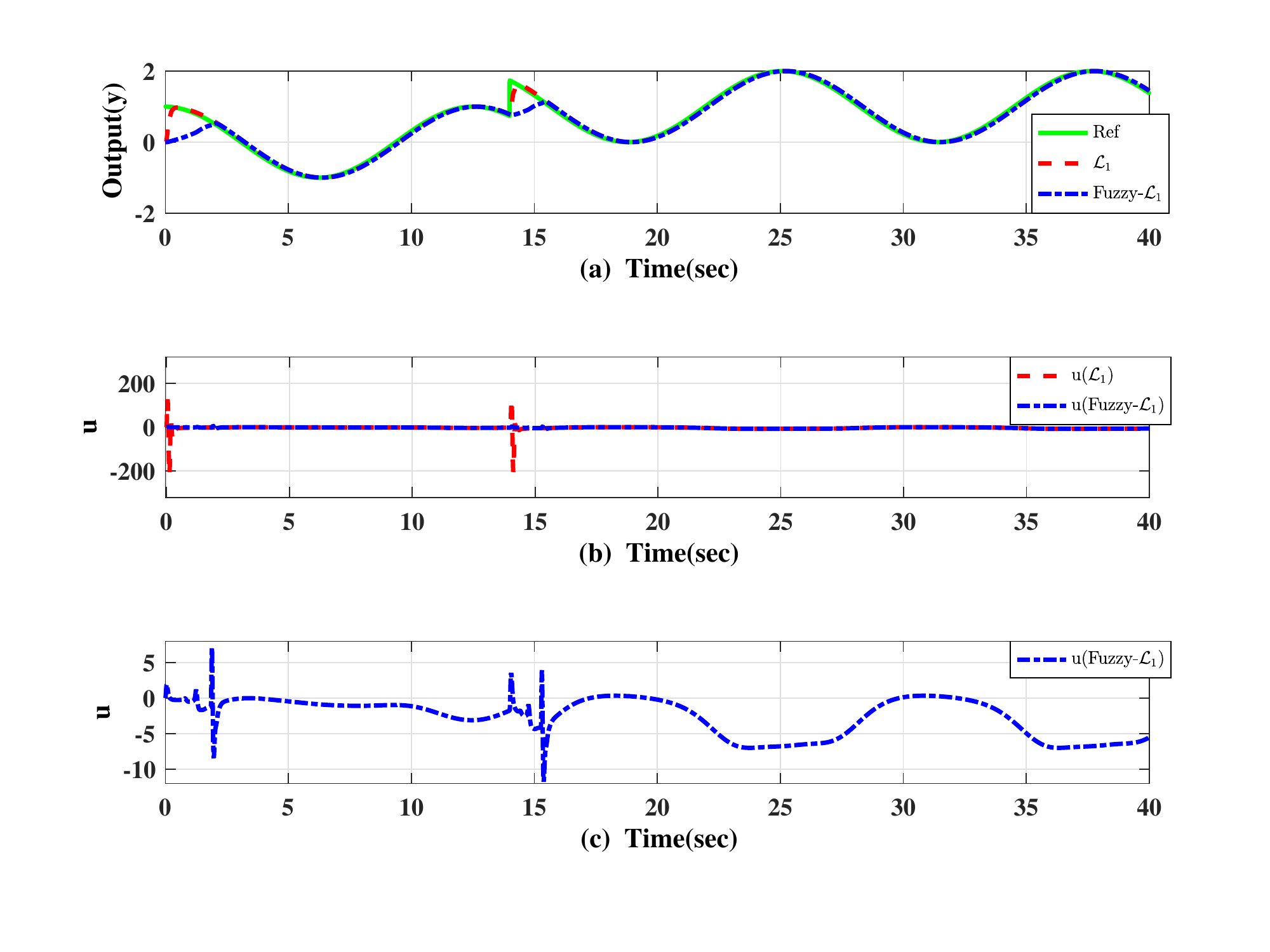}
		\caption{Performance of fuzzy-\Lone adaptive controller and \Lone adaptive controller for nonlinear system of case 1.}\label{Fuzzy_L1_out2}
	\end{figure*}
	
	\begin{figure*}[ht]
		\centering
		\includegraphics[scale=0.5]{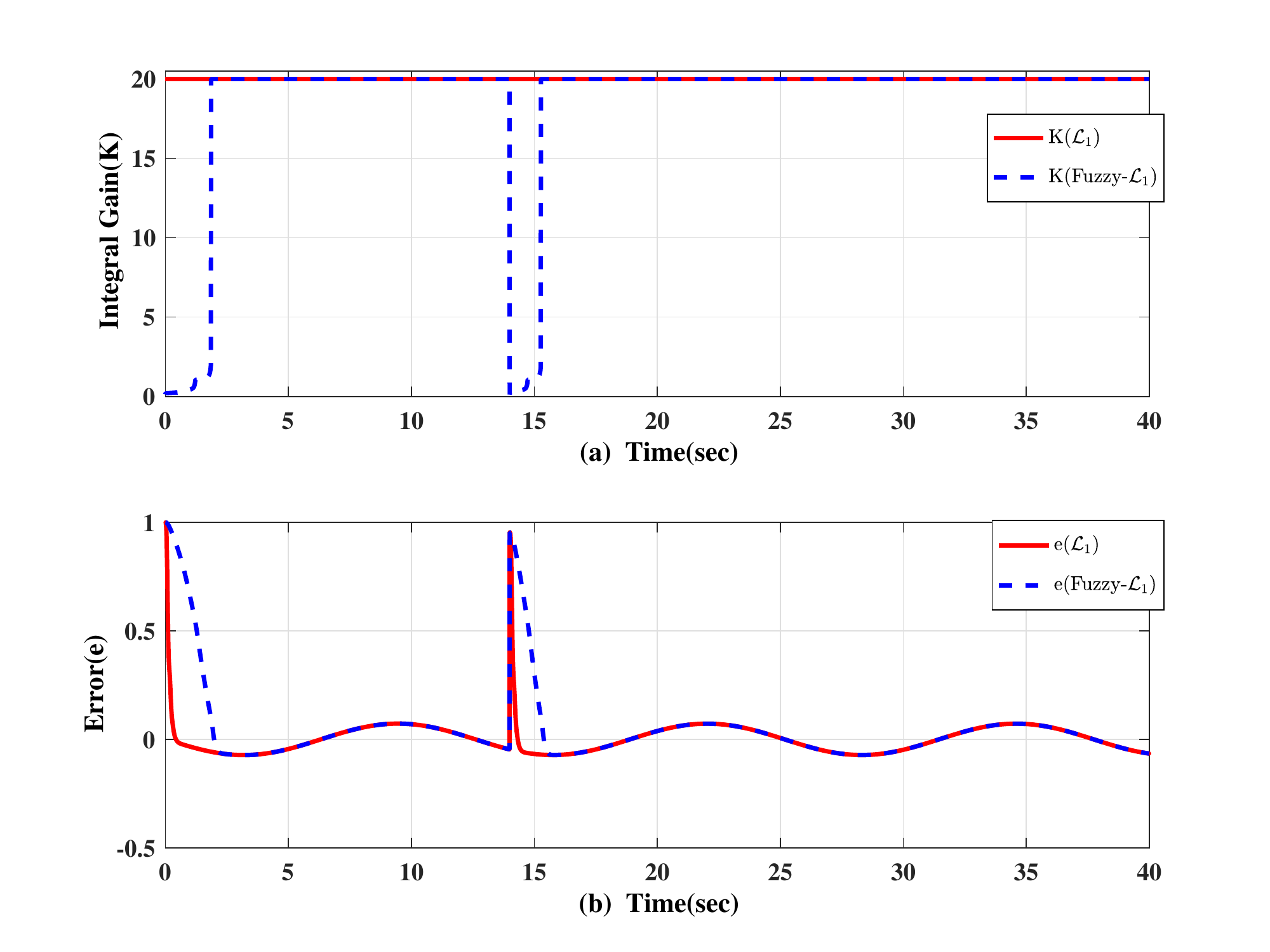}
		\caption{Feedback gain and output error of fuzzy-\Lone adaptive controller and \Lone adaptive controller of case 1.}\label{Fuzzy_L1_out2_2}
	\end{figure*}
	{\bf Case 2:}To illustrate the effectiveness of the proposed fuzzy filter with \Lone adaptive controller, robustness of fuzzy filter is examined against any existing of high uncertainties, unmodeled input parameters and disturbances.Here, the nonlinear model and other assumptions mentioned in case 1 are similar the nonlinear function; however, the nonlinearity includes high time variant uncertainties and disturbances and these changes will except be presented as follows
	\begin{equation*}
	\begin{split}
	f&(x(t),t) = \big(sin(0.4t)+1 \big)x_1^2(t) + \big(2cos(0.35t)+0.5 \big)x_2^2(t)\\
	&+ \big(sin(0.3t)+0.3 \big)x_1sin(x_1^2) + sin(0.35t)cos(0.4t)\\ 
	& + 0.5x_2cos(x_2^2+0.5cos(0.3t))+ sin(0.3t)cos(0.4t)z^2 
	\end{split}
	\end{equation*}
	where
	\begin{equation*}
	z(s) = \frac{s-1}{s^2+3s+2}v(s),\hspace{10pt} v(t) = x_1sin(0.2t)+x_2\\    
	\end{equation*}
	The robustness of fuzzy feedback filter gain with \Lone adaptive controller has been validated in Figure \ref{Fuzzy_L1_out3} and presented versus \Lone adaptive controller. The significant impact and the advantage of fuzzy-\Lone controller is revealed on control signals performance as shown Figure \ref{Fuzzy_L1_out3}. Figure \ref{Fuzzy_L1_out3_2}.(a) presents the performance of feedback gain for fuzzy-\Lone adaptive controller and \Lone adaptive controller. Finally, Figure \ref{Fuzzy_L1_out3_2}.(b) shows the error of both controllers. Uniform transient and tracking capability are validated as shown in Figure \ref{Fuzzy_L1_out2}  and \ref{Fuzzy_L1_out3}. The benefits of fuzzy-\Lone adaptive controller can be summarized in including fast desired dynamics and improving the tracking capability and robustness with less range of control signal.
	\begin{figure*}[ht]
		\centering
		\includegraphics[scale=0.5]{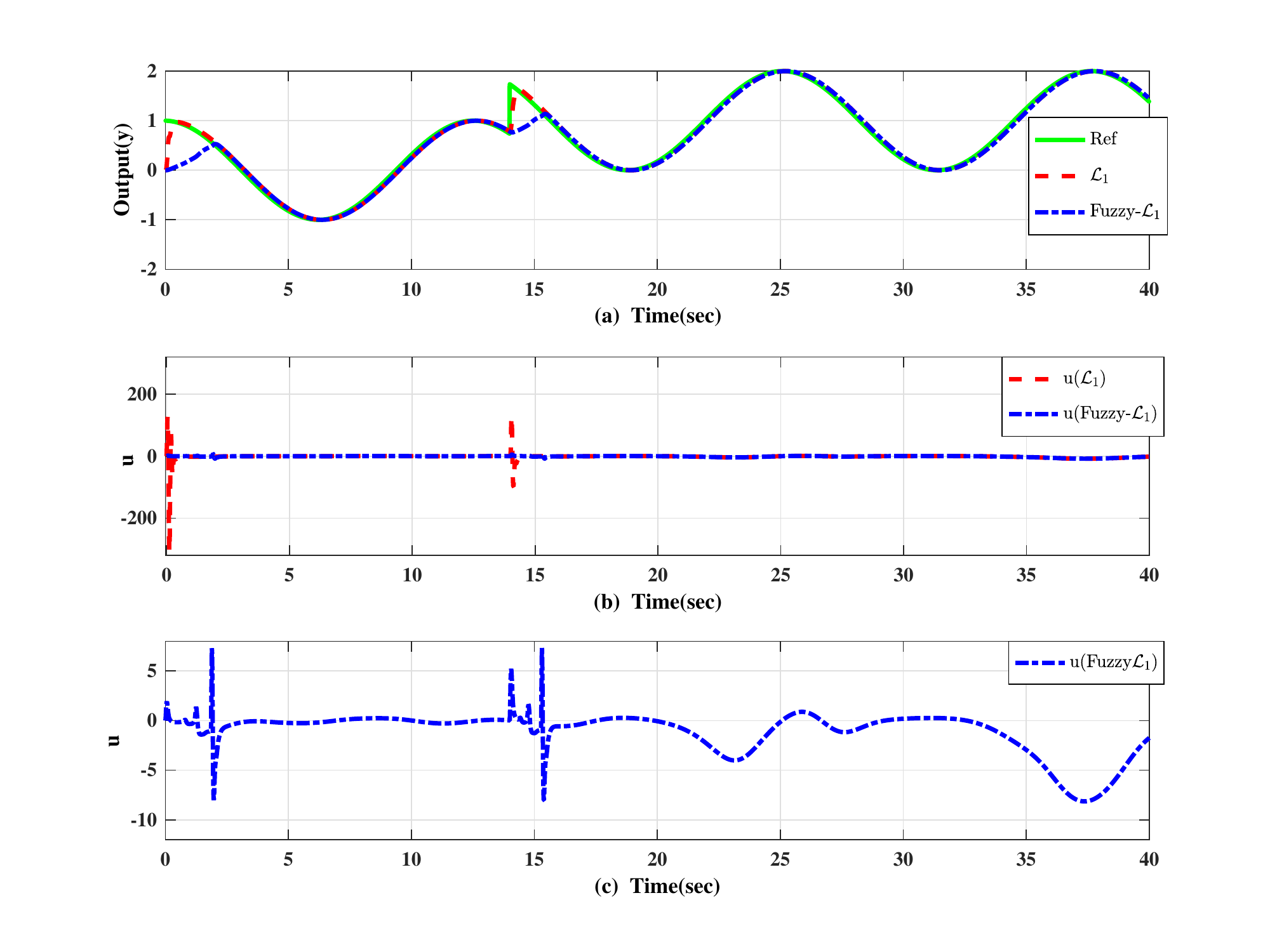}
		\caption{Performance of fuzzy-\Lone adaptive controller and \Lone adaptive controller for nonlinear system of case 2.}\label{Fuzzy_L1_out3}
	\end{figure*}
	
	\begin{figure*}[ht]
		\centering
		\includegraphics[scale=0.5]{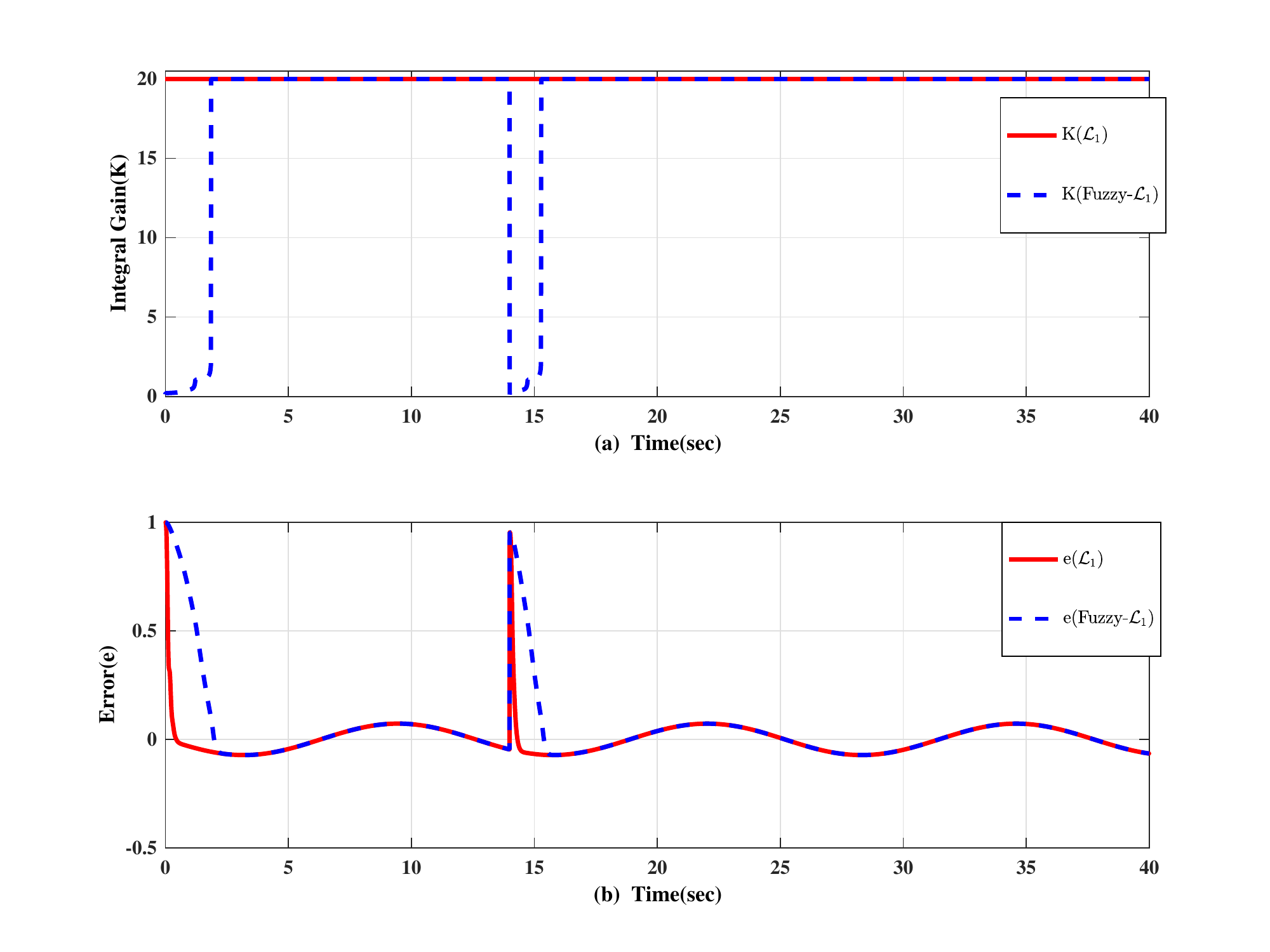}
		\caption{Feedback gain and output error of fuzzy-\Lone adaptive controller and \Lone adaptive controller of case 2.}\label{Fuzzy_L1_out3_2}
	\end{figure*}
	
	{\bf Case 3:} The robustness of fuzzy-\Lone adaptive controller and \Lone adaptive controller will reveal more in this case. All aforementioned assumptions in case 2 are similar here except the desired closed loop dynamics assumed to be faster than case 2. Desired poles are set to $p=-84 \pm j0.743$. According to this change in closed loop poles, the robustness of \Lone adaptive controller will be violated and the system will no longer be stable. However, fuzzy-\Lone adaptive controller will be able to track the output under this new condition with limitation in increasing the control signal range. Figure \ref{Fuzzy_L1_out4} illustrate the output performance of fuzzy-\Lone adaptive controller for case 3.\\
	\begin{figure*}[ht]
		\centering
		\includegraphics[scale=0.6]{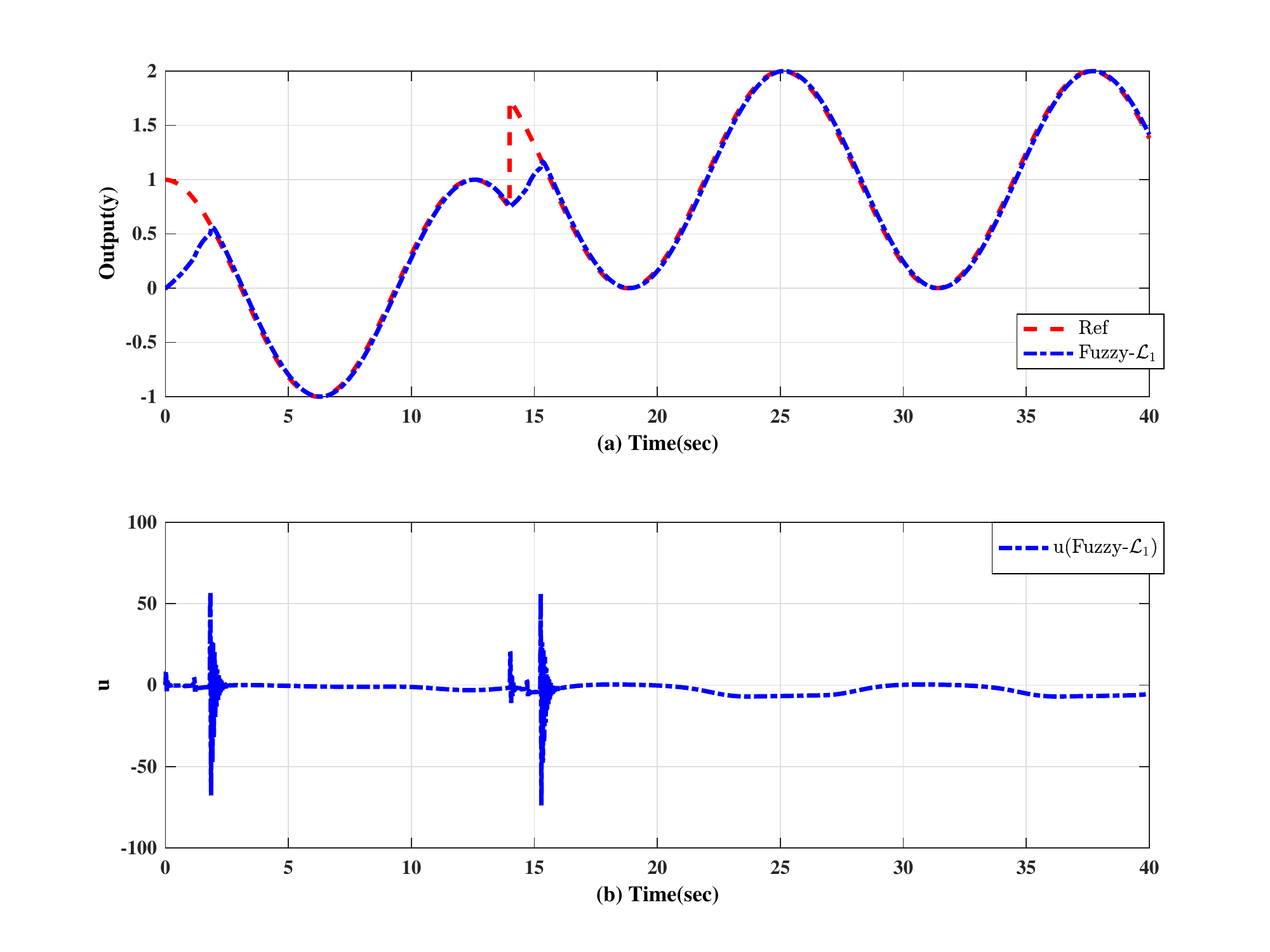}
		\caption{Performance of fuzzy-\Lone adaptive controller for nonlinear system of case 3.}\label{Fuzzy_L1_out4}
	\end{figure*}
	The robustness of this criterion has been simulated and validated with \Lone adaptive controller on high nonlinear system with different forms of nonlinearities and uncertainties in addition to fast closed loop dynamics compared to normal structure of \Lone adaptive controller.  It can be concluded based on the cases considered and results obtained that the proposed fuzzy-based approach to tune the feedback filter improves greatly the performance of \Lone adaptive controller. The proposed fuzzy-\Lone adaptive controller guarantee boundedness of the output and control signal and insures fast tracking and low range of control signal.
	
	\section{Conclusion} \label{Sec_6}
	This paper presents a new FLC-PSO design of the feedback gain filter part of  \Lone adaptive controller. PSO determines the optimal variables of the output membership functions. The proposed algorithm tunes on-line the filter parameters, which in turn contributed to improving robustness and stability of \Lone adaptive controller. Moreover, owing to a smooth tuning of the filter the control signal range has been greatly reduced. Illustrative examples were developed and simulated to compare fuzzy-\Lone adaptive controller with \Lone adaptive controller with constant filter parameters and to validate the advantages of the proposed approach. The results show improved performance and robustness with high levels of time variant uncertainties and disturbances in addition to fast desired closed loop adaptation. 
	
	There are several directions for future work. One important area is to implement this approach on a real system and compare the performance with existing techniques. One should note that computation power is relatively cheap and the technology offers several hardware option over which this controller can be implemented. This study aimed at proposing an effective way of tuning the coefficients of the control filter. During this work, the structure of the filter was fixed. Extending this work to determine automatically the appropriate filter's structure has the potential to further improve the robustness and stability. Such an extension will take into account the health of the system and may lead to the design a failure tolerant robust controller.  In this study, PSO has been implemented off-line. To our knowledge, recursive PSO for online implementation is not explored. A recursive and online PSO will impose hard constraints on hardware capacity. The implementation of PSO as it is now is not feasible for online control implementation.  A comparison between PSO and other evolutionary algorithms can be established to define the most effective solution for the fuzzy-\Lone adaptive control problem. 
	
	\section*{Acknowledgment}
	The author(s) would like to acknowledge the support of Deanship of Scientific Research, King Fahd University of Petroleum and Minerals, through the Electrical and Energy System Research group funded project \#RG1116-1\&2.
	
	In addition, The authors would like to acknowledge the support of Deanship of Scientific Research, King Fahd University of Petroleum and Minerals, through the Electrical and Energy System Research group funded project \# RG1207-1.

	\bibliographystyle{apacite}
	\bibliography{Bib_L1_Fuzzy}

\end{document}